\documentclass[fleqn,twoside,twocolumn,nofootinbib,showkeys]{revtex4} 
\usepackage[sec,nocpr]{ujp} 

\begin{document}
\title[Resonant Threshold Two-Photon $e^{-}e^{+}$ Pair]%
{RESONANT THRESHOLD TWO-PHOTON \boldmath$e^{-}e^{+}$\\ PAIR
PRODUCTION ONTO THE LOWEST
LANDAU\\ LEVELS IN A STRONG MAGNETIC FIELD}%
\author{M.M.~Dyachenko}
\affiliation{Institute of Applied Physics, Nat. Acad. of Sci. of
Ukraine}
\address{58, Petropavlivska Str., Sumy 40000, Ukraine}
\email{dia4enkomisha@yandex.ru,\\ novak-o-p@ukr.net,
kholodovroman@yahoo.com}
\author{O.P.~Novak}
\affiliation{Institute of Applied Physics, Nat. Acad. of Sci. of
Ukraine}
\address{58, Petropavlivska Str., Sumy 40000, Ukraine}
\author{R.I.~Kholodov}
\affiliation{Institute of Applied Physics, Nat. Acad. of Sci. of
Ukraine}
\address{58, Petropavlivska Str., Sumy 40000, Ukraine}

\udk{539.1.01} \pacs{12.20.-m, 13.88.+e} \razd{\seci}

\autorcol{M.M.\hspace*{0.7mm}Dyachenko, O.P.\hspace*{0.7mm}Novak,
R.I.\hspace*{0.7mm}Kholodov}


\setcounter{page}{849}%

\begin{abstract}
The process of electron-positron pair production by two photons in a
strong magnetic field has been studied.\,\,The process kinematics is
considered, and the probability amplitude for arbitrary
polarizations of particles is found.\,\,The resonance conditions are
established, and the resonant cross-section is estimated in the case
where the electron and the positron occupy the lowest levels
($l_{e}=$~1, $l_{p}=$~0) that satisfy these resonance conditions.
\end{abstract}

\keywords{electron-positron pair, photon-induced production,
magnetic field, quantum electrodynamics, resonance.}

\maketitle

\section{Introduction}\vspace*{-0.5mm}

The researches of quantum electrodynamic (QED) processes that take
place at the collisions of heavy ions have a challenging character,
which is associated with the progress achieved in the domain of
heavy-ion accelerators.\,\,In such experiments, the main attention
is usually concentrated on the researches dealing with properties of
the strong interaction.\,\,However, the QED processes can also be
studied even if nuclei do not approach each other at the distance
characteristic of the strong interaction \cite{Baur02, Baur07}.

In the course of collisions between heavy ions, the extremely
intensive, rapidly varying electromagnetic fields are
generated.\,\,The resulting strength of the field can exceed the
critical QED value $H_{c}=\frac{m^{2}c^{3}}{e\hbar }\approx$
$\approx4.41\times10^{13}$~Gs, which, according to the theoretical
reasoning, is the vacuum excitation threshold
\cite{Greiner}.\,\,This fact opens vast opportunities for testing
the QED predictions in the range of strong fields.\,\,A wide program
of such experiments is planned to be fulfilled on the future FAIR
(Facility for Antiproton and Ion Research) installation, which is
under construction on the basis of the GSI (GSI Helmholtz Centre for
Heavy Ion Research, Darmstadt, Germany) \cite{CDR}.

The production of electron-positron pairs at the collisions between
ions is a challenging problem from the theoretical and experimental
viewpoints.\,\,In particular, the production of a pair with the
electron in a bound state changes the charge of one of the ions and
is one of the principal origins of beam losses in modern colliders.
On the other hand, the pair production by a strong total
supercritical field of heavy nuclei is the evidence of a bound state
diving in the negative energy continuum, as well as of the
transition of the vacuum into a new state \cite{Greiner}.

A poorly studied issue in the problem of $e^{-}e^{+}$ pair
production at the collision of moving ions is the influence of the
magnetic field created by the ions.\,\,Simple estimates testify that
the field between the ions can reach the critical value $H_{c}$ or
can even exceed it already at collision energies of the Coulomb
barrier order.\,\,However, in the majority of researches, only the
Coulomb field of ions is taken into account.\,\,The attempts to
reveal the influence of a magnetic field were made for the first
time in works \cite{Soff81, Rumrich87, Soff88} in the framework of
the model of quasimolecule (for low-energy collisions).\,\,The
corresponding numerical calculations showed the absence of the
effect; however, the interaction of the created pair with the
magnetic field was not taken into consideration.

In work \cite{Fomin98}, the assumption that the created
elect\-ron-po\-si\-tron pair captures the magnetic field similarly
to the phenomenon, where the magnetic field lines \textquotedblleft
are frozen\textquotedblright\ into a plasma, was
substantiated.\,\,The corresponding estimates testify that the
lifetime of the self-consistent system \textquotedblleft pair +
magnetic field\textquotedblright\ considerably exceeds the nuclear
transit time.\,\,Therefore, the magnetic field can substantially
affect the process.\,\,The presence of characteristic resonances in
the corresponding channels can be the observable evidence of QED
processes in the magnetic field created by colliding ions.

It has to be noted that, in the EPOS and ORANGE experiments carried
out at the GSI, anomalous peaks were revealed in the channel of
electron-positron pair production \cite{Backe78, Koenig87,
Cowan86}.\,\,The nature of peaks was not elucidated, and their very
existence was not confirmed.\,\,In view of the similarity between
the spectrum of anomalous peaks and the quasiequidistant spectrum of
electron energy levels in the magnetic field, it is quite probable
that further researches of the role of a magnetic field in ionic
collisions will deepen our understanding of this phenomenon.

According to the results of work \cite{Fomin98}, the lifetime of the
magnetic field substantially exceeds the characteristic time of
electromagnetic in\-te\-rac\-ti\-on.\,\,The\-re\-fo\-re, the process
can be considered as the electron-positron pair production in a
constant uniform magnetic field.\,\,This approximation allows one to
reveal the main features of the process and, at the same time, to
obtain simple analytical expressions for the
cross-section.\,\,Moreover, if the energy of ions is sufficiently
high, the conditions for the application of the equivalent-photon
method \cite{LandauIV} are satisfied.\,\,Hence, in the simplest
case, our problem is approximately reduced to the study of the
process of two-photon electron-positron pair production in a
magnetic field.

Let us also mention possible applications of the results of this
work to the problem of electron cooling of heavy-ion beams.\,\,The
essence of the method consists in a decrease of the effective beam
temperature owing to the collision of ions in the beam with
electrons that move together with the beam and possess a small
spread of velocities \cite{Budker78, Meshkov94, Parkhomchuk00}.
Although the method of electron cooling has been known for a long
time and is widely used in accelerating facilities, the
experimentally observed difference between the cooling efficiency
for positive and negative particles remains unexplained
\cite{Dikanskii88}.\,\,In particular, the electron cooling of an
antiproton beam is planned to be used in the HESR accumulator
(FAIR), and this fact substantiates the problem urgency
\cite{Galnander09, Bazhenov03}.

The application of the method of quantum field theory to the problem
of electron cooling \cite{Larkin59, Akhiezer61} has considerable
advantages in comparison with others (the method of pair collisions,
the dielectric model, and so forth), because it contains no
phenomenological parameters.\,\,The corresponding stopping power is
expressed in terms of the polarization operator.\,\,The difference
between the cooling of protons and antiprotons may probably be
described in the second Born approximation.\,\,However, according to
the optical theorem, the imaginary part of the polarization operator
in the second order is equal to the total probability of the process
of two-photon pair production.

For the first time, the process of two-photon
elect\-ron-po\-si\-tron pair production in a magnetic field was
considered in work \cite{Ng77} in the simplest case of the head-on
photon collision along the magnetic field.\,\,In work
\cite{Kozlenkov86}, this process was analyzed for an arbitrary
direction of photon propagation, but under the assumption that the
energy of each photon is insufficient for the pair production in the
one-photon process.\,\,This condition excludes the possibility of
the resonance character of the reaction.

\begin{figure*}
  \includegraphics[width =12cm]{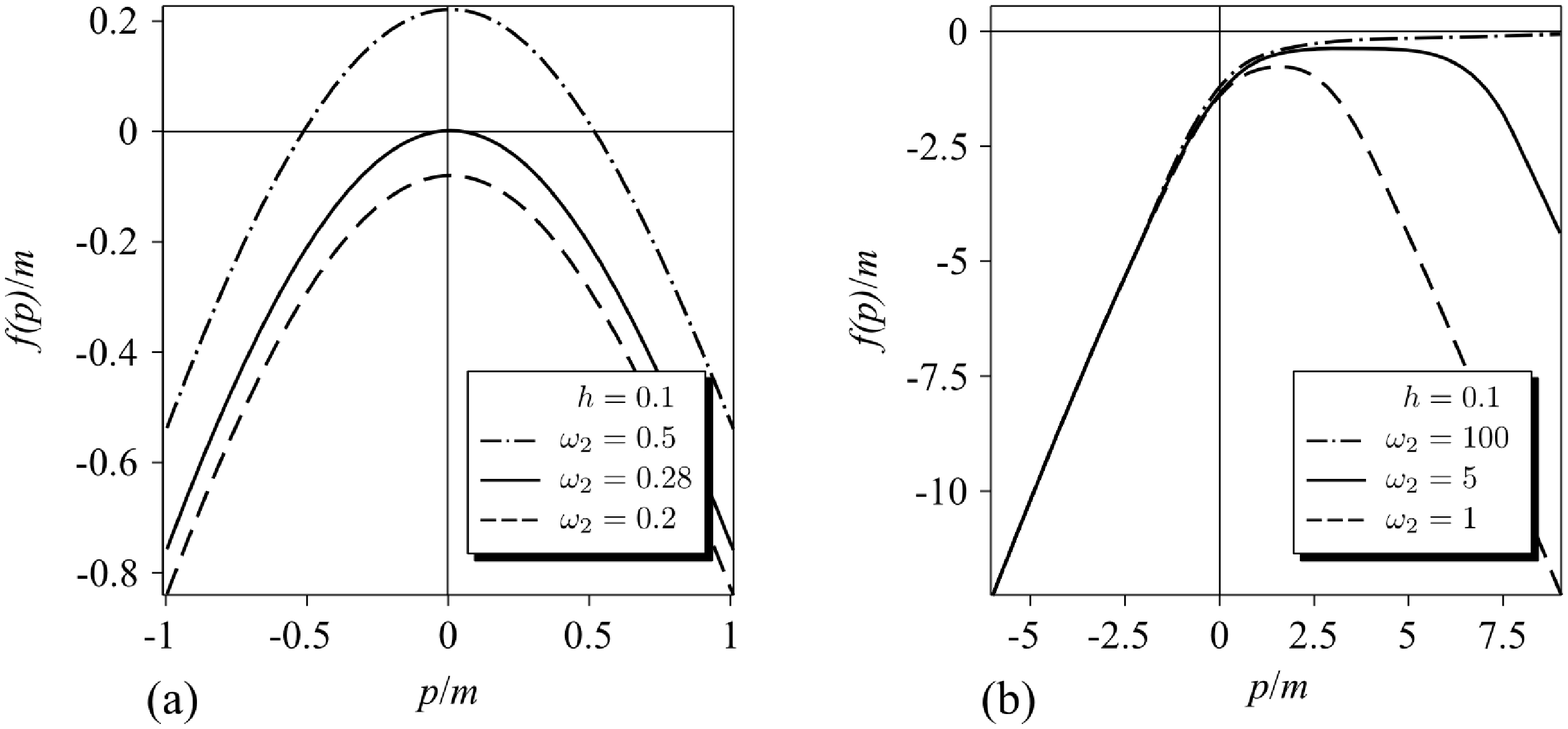}
\vskip-3mm \parbox{12cm}{\caption{Function $f(p_{ez})/m$ plotted for
the numbers of Landau levels $l_{e}=2$ and $l_{p}=1$.\,\,The
frequency of the first photon in terms of the electron mass units,
$\omega_{1}=2$. $\theta_{1,2}=\pi/2$ (\textit{a}) and 0 (\textit{b})
}}
\end{figure*}

In this work, in contrast to the earlier researches, the process was
considered in the case where the energy of one of the photons or
each of them is sufficient for the pair production.\,\,The total
amplitude of the process with arbitrary polarizations of particles
and the conditions for the resonant character of the process are
determined.\,\,The process cross-section is estimated for the lowest
Landau levels that satisfy the resonance conditions.

Throughout the work, the relativistic system of units, $\hbar=c=1$, is used.

\section{Kinematics}

It is known \cite{LandauIV} that the electron occupies discrete energy levels
in a uniform magnetic field,
\begin{equation}
  E = \sqrt{p_z^2 + \tilde m^2}, \quad
  \tilde{m} = m\sqrt{1 + 2 l h},
\end{equation}
where $p_{z}$ is the longitudinal momentum component with respect to
the field, $l$ the level number, and $h$ the magnetic field measured
in the units of $H_{c}$.\,\,In the magnetic field, the conservation
laws of the energy and the momentum component parallel to the field
are obeyed and have the following form for the process of two-photon
electron-positron pair production:\vspace*{-1mm}
\begin{equation}
  \label{laws}
    \begin{array}{l}
      \displaystyle E_e + E_p = W, \\[1mm]
      \displaystyle p_{ez} + p_{pz} = K.
    \end{array}
\end{equation}
Here, $W=\omega_{1}+\omega_{2}$ is the total energy, and
$K=k_{1z}\,+$ $+\,k_{2z}$ is the total longitudinal momentum of
photons.

Let us determine the threshold conditions of the pair production
onto some Landau levels $l_{e}$ and $l_{p}$.\,\,First, let us
introduce the auxiliary function
$f(p_{ez})=\omega_{1}+\omega_{2}-E_{e}-E_{p}$.\,\,The process
threshold is determined by the condition $f(p_{m})=0$, where $p_{m}$
is the point of a maximum of the function.\,\,After the
differentiation, we find the value of electron momentum at the
maximum,\vspace*{-1mm}
\begin{equation}
  \label{maximum}
  p_m = \frac{\tilde m_e K}{\tilde m_e + \tilde m_p}.
\end{equation}
In view of the previous relations, the threshold values of the
energy and the electron and positron momenta are expressed as
follows:\vspace*{-1mm}
\begin{equation}
  \label{E_th}
    E_e' = \frac{\tilde m_e}{\tilde m_e + \tilde m_p} W', \quad
    p_{ez}' = \frac{K'E_e'}{W'},
\end{equation}\vspace*{-6mm}
\begin{equation}
    E_{p}' = \frac{\tilde m_p}{\tilde m_e + \tilde m_p} W', \quad
    p_{pz}' = \frac{K'E_p'}{W'}.
\end{equation}
From whence, the condition for the reaction threshold takes the
form\vspace*{-1mm}
\begin{equation}
  \label{threshold}
  W'^2 - K'^2 = (\tilde m_e + \tilde m_p)^2.
\end{equation}

In the general case (above the threshold), system (\ref{laws}) gives
the following expression for the momenta of particles:\vspace*{-2mm}
\begin{equation}
\label{pz}
  p_{ez} = \frac{aK \pm bW}{2M^2}.
\end{equation}
Here, the notations\vspace*{-1mm}
\begin{equation}
    a = M^2 - \tilde{m}_p^2 + \tilde{m}_e^2\!,
 \end{equation}\vspace*{-9mm}
\begin{equation}
    b^2 = M^4 -2M^2((\tilde m_e)^2 + (\tilde m_p)^2) + 4(l_p - l_e)^2h^2  \!,
\end{equation}\vspace*{-9mm}
\begin{equation}
    M^2 = W^2 -K^2
\end{equation}\vspace*{-5mm}

\noindent were introduced.\,\,The process is evidently impossible,
if both photons move in parallel to ${\bf H}$, because $K^{2}=W^{2}$
in this case, and condition (\ref{threshold}) is not
satisfied.\,\,This situation is illustrated in Fig.~1.

Note that the Lorentz transformations along ${\bf H}$ do not change
the magnetic field itself.\,\,Therefore, without any loss of
generality, the reference frame can be chosen to exclude the
longitudinal momentum of photons: $K=0$.\,\,In this reference frame,
$p_{ez}=-p_{pz}$, and the threshold conditions
have a simpler form,\vspace*{-1mm}%
\begin{equation}
  W' = \tilde{m}_e + \tilde{m}_p, \quad p_{ez}' = p_{pz}' = 0.
\end{equation}

\section{Amplitude}

The electron and positron wave functions look like\cite{Fomin00}\vspace*{-3mm}
\begin{equation}
\label{Psi}
  \begin{array}{l}
\displaystyle \Psi_e = \frac{1}{\sqrt{S}}e^{-i(Et - p_y y - p_z z)}
\psi_e(\zeta^-),
    \\[4mm]
\displaystyle \Psi_p = \frac{1}{\sqrt{S}}e^{+i(Et - p_y y - p_z z)}
\psi_p(\zeta^+),
  \end{array}
\end{equation}
where $\zeta^{\pm}=m\sqrt{h}(x\mp p_{y}/m^{2}h)$,%
\[
  \psi_e (\zeta^-) = C_e
  \biggl[
  i\sqrt{\tilde{m}_e - \mu_e m}  U_l(\zeta^-) \,+
\]\vspace*{-7mm}
\begin{equation}\label{psie}
  + \,\mu_e\sqrt{\tilde{m}_e + \mu_e m}U_{l-1}(\zeta^-)\gamma^1
  \biggr]u_e ,
\end{equation}\vspace*{-7mm}
\[
  \psi_p (\zeta^+) = C_p
  \Bigl[
  i\sqrt{\tilde{m}_p + \mu_p m}  U_l(\zeta^+) \,-
  \]\vspace*{-7mm}
\begin{equation}\label{psip}
  -\, \mu_p\sqrt{\tilde{m}_p - \mu_p m}U_{l-1}(\zeta^+)\gamma^1
  \Bigr]u_p .
\end{equation}
Here, $S$ is the normalization area, $U_{l}(x)$ the Hermite function,
$\mu_{e,p}$ are the doubled spin projections of the electron and the positron,
$C_{e,p}$ the normalization constants, and $u_{e,p}$ the constant bispinors:%
\begin{equation}
  C_{e,p} = \frac 12 \sqrt{\frac{\sqrt{eH}}{E_{e,p}\tilde{m}_{e,p}}}\!,
\end{equation}\vspace*{-7mm}
  \begin{equation}
  u_{e,p} = \frac{1}{R_{e,p}} \left(0,\: \pm R^2_{e,p},\: 0,\: p_{{(e,p)} z}\right)\!,
\end{equation}\vspace*{-7mm}
  \begin{equation} \label{Rep}
  R_{e,p} = \sqrt{E_{e,p} - \mu_{e,p} \tilde m_{e,p}}, 
\end{equation}
where the sign \textquotedblleft$-$\textquotedblright\ in $u_{e,p}$
corresponds to the elect\-ron.\,\,The wave functions
(\ref{Psi})--(\ref{psip}) correspond to the $(0;0,xH,0)$ gauge of
the electromagnetic 4-potential.

\begin{figure}
\vskip-2mm
  \includegraphics[width = \columnwidth]{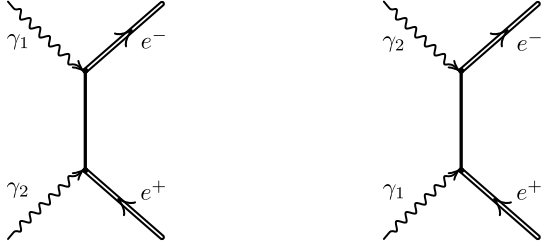}
\vskip-5mm  \caption{Feynman diagrams for the two-photon $e^{-}e^{+}$
pair production in a magnetic field  }
\end{figure}

For the wave function of the initial photon, the standard expression is used
\cite{LandauIV},%
\begin{equation}
  A^\nu = \sqrt{\frac{2\pi}{\omega V}} e^\nu e^{ikx}\!,
\end{equation}
where $V$ is the normalization volume, $e^\nu = (0, {\bf e})$ is the
4-vector
of photon polarization, and%
\begin{equation}
  \mathrm{e} = \left(\!\!
  \begin{array}{c}
    \cos\phi \cos\theta \cos\alpha - e^{i\beta} \sin\phi \sin\alpha   \\
    \sin\phi \cos\theta \cos\alpha + e^{i\beta} \cos\phi \sin\alpha   \\
    - \sin\theta \cos\alpha
  \end{array}
 \!\! \right)\!\!.
\end{equation}
Here, $\phi$ and $\theta$ are the azimuthal and polar angles, respectively; and
$\alpha$ and $\beta$ are polarization parameters.

Green's function of a virtual electron looks like
\cite{Fomin99}\vspace*{-2mm}
\begin{equation}
\label{G}
  G(x-x') = \frac{-m\sqrt{h}}{(2\pi)^3} \int d^3g e^{-i\Phi} \sum\limits_{n} \frac{G_H(x,x')}{g_0^2 -
  E_n^2},
\end{equation}\vspace*{-8mm}
\[
G_H(x,x') = U_n U_n' (\gamma P + m) \tau\, +
\]\vspace*{-9mm}
\[
    +\, im \sqrt{2nh} U_{n-1}U'_{n}\gamma^1\tau - im \sqrt{2nh} U_n U_{n-1}' \tau
    \gamma^1\,+
\]\vspace*{-9mm}
\begin{equation}\label{GH}
    +\, U_{n-1} U'_{n-1} (\gamma P + m)\tau^*,
\end{equation}\vspace*{-5mm}

\noindent where $\gamma$'s are Dirac gamma matrices,\vspace*{-1mm}
\begin{equation}
  \Phi = g_0(t-t') - g_y (y-y') - g_z (z-z'),
\end{equation}\vspace*{-8mm}
\begin{equation}
  \tau = \frac{1}{2} (1 + i \gamma_2\gamma_1),
\end{equation}\vspace*{-8mm}
\begin{equation}
  P = (E_n, 0, 0, g_z).
\end{equation}\vspace*{-5mm}

\noindent The argument of Hermite functions in Eq.~(\ref{GH}) looks
like\vspace*{-2mm}
\[
  \rho(x) = m\sqrt{h} x + g_y/m\sqrt{h},
\]\vspace*{-5mm}

\noindent and the primed function means its dependence on
$\rho(x^{\prime})$.

According to the QED rules, the two-photon electron-positron pair
production amplitude is determined as \vspace*{-1mm}
\[
  S_{fi} = -ie^2 \int d^4x d^4x'
  \Bigl[
  \bar\Psi_e (A_1\gamma) G(x-x') (A'_2\gamma) \Psi_p' \,+
 \]\vspace*{-7mm}
 \begin{equation}\label{Sfi}
   +\, \bar\Psi_e (A_2\gamma) G(x-x') (A'_1\gamma) \Psi_p'
  \Bigr],
\end{equation}\vspace*{-5mm}

\noindent where the primed quantities depend on
$x^{\prime}$.\,\,Figure~2 demonstrates the Feynman diagrams that
correspond to amplitude (\ref{Sfi}).\,\,Substituting the wave
functions and the propagator into Eq.\,(\ref{Sfi}) and carrying out
the corresponding transformations, we obtain the amplitude in the
form\vspace*{-2mm}
\[
     S_{fi}^{(1)} = \frac{-ie^2  (2\pi)^4 }
    {4VS \sqrt{\omega_1\omega_2\tilde{m}_e\tilde{m}_p E_e E_p}}
    \sum\limits_{n=0}^{\infty}
    \frac{\sum_{j=1}^{16} B_{j}}{g_0^2 - E_n^2}\,
    \times
\]\vspace*{-7mm}
\begin{equation}
\label{Sfi1}
    \times\,
    \delta^3(\gamma_1 + \gamma_2  - e^- - e^+),
\end{equation}
\vspace*{-5mm}

\noindent where the terms $B_{j}$ look like\vspace*{-2mm}
\[
  B_{1} =
  -i\mu_e M_{e+}^{p+}  e_1T_2^-
  \,\times
\]\vspace*{-9mm}
\[
   \times\,
  (mS^+ \! + \! E_n S^- \!\! - \! g_z A^-)
  I^{(1)}_{l_e - 1, n -1} I^{(2)}_{n - 1, l_p},
\]\vspace*{-9mm}
\[
  B_{2} =
  \mu_e \mu_p M_{e+}^{p-}  e_{1}e_{2}
  \,\times
\]\vspace*{-9mm}
\[
  \times\,
  (mA^- \! + E_n A^+ \! + g_z S^+)
  I^{(1)}_{l_e - 1, n -1} I^{(2)}_{n - 1, l_p - 1},
\]\vspace*{-9mm}
\[
  B_{3} =
  i\mu_e M_{e+}^{p+}  e_2T_1^-
  (mS^+ \! \! + E_n S^- \! \! + \! g_z A^-)
  I^{(1)}_{l_e - 1, n} I^{(2)}_{n, l_p},
\]\vspace*{-9mm}
\[
  B_{4} =
  \mu_e \mu_p M_{e+}^{p-}  T_1^- T_2^+
  \,\times
\]\vspace*{-9mm}
\[
  \times\,
  (mA^- \! + E_n A^+ \! - g_z S^+)
  I^{(1)}_{l_e - 1, n} I^{(2)}_{n, l_p - 1},
\]
\[
  B_{5} =
  - \mu_e M_{e+}^{p+} m\sqrt{2nh} \:  e_1e_2 A^-
  I^{(1)}_{l_e - 1, n - 1} I^{(2)}_{n, l_p},
\]\vspace*{-7mm}
\[
  B_{6} =
  -i \mu_e \mu_p M_{e+}^{p-} m\sqrt{2nh} \:  e_1 T_2^+  S^+
  I^{(1)}_{l_e - 1, n - 1} I^{(2)}_{n, l_p - 1},
\]\vspace*{-7mm}
\[
  B_{7} =
  - \mu_e M_{e+}^{p+} m\sqrt{2nh} \:  T_1^- T_2^- A^-
  I^{(1)}_{l_e - 1, n } I^{(2)}_{n -1 , l_p},
\]\vspace*{-7mm}
\[
  B_{8} =
  -i \mu_e \mu_p  M_{e+}^{p-} m\sqrt{2nh} \:  e_2 T_1^- S^+
  I^{(1)}_{l_e - 1, n } I^{(2)}_{n -1 , l_p - 1},
\]\vspace*{-7mm}
\[
  B_{9} =
  M_{e-}^{p+}  T_1^- T_2^+
  \,\times
\]\vspace*{-7mm}
\[
  \times\,
  (mA^- \! \! - E_n A^+ \! \! + g_z S^+)
  I^{(1)}_{l_e, n - 1} I^{(2)}_{n - 1, l_p},
\]\vspace*{-7mm}
\[
  B_{10} =
  -i\mu_p M_{e-}^{p-} e_2 T_1^+
 \, \times
\]\vspace*{-7mm}
\[
  \times\,
  (mS^+ \!\! - E_n S^- \!\! - g_z A^-)
  I^{(1)}_{l_e, n - 1} I^{(2)}_{n - 1, l_p - 1},
\]\vspace*{-7mm}
\[
  B_{11} =
  M_{e-}^{p+}  e_1e_2
  (mA^- \!\! - E_n A^+ \!\! - g_z S^+)
  I^{(1)}_{l_e, n} I^{(2)}_{n, l_p},
\]\vspace*{-7mm}
\[
  B_{12} =
  i \mu_p M_{e-}^{p-} e_1 T_2^+
  \,\times
\]\vspace*{-7mm}
\[
   \times\,
  (m S^+ \!\! - E_n S^- \!\! + g_z A^-)
  I^{(1)}_{l_e, n} I^{(2)}_{n, l_p - 1},
\]\vspace*{-7mm}
\[
  B_{13} =
  -i M_{e-}^{p+} m\sqrt{2nh} \:  e_2 T_1^+ S^+
  I^{(1)}_{l_e, n - 1} I^{(2)}_{n, l_p},
\]\vspace*{-7mm}
\[
  B_{14} =
  \mu_p M_{e-}^{p-} m\sqrt{2nh} \:  T_1^+ T_2^+ A^-
  I^{(1)}_{l_e, n - 1} I^{(2)}_{n, l_p -1},
\]\vspace*{-7mm}
\[
  B_{15} =
  -i M_{e-}^{p+} m\sqrt{2nh} \:  e_1 T_2^- S^+
  I^{(1)}_{l_e, n} I^{(2)}_{n - 1, l_p},
\]\vspace*{-7mm}
\[
  B_{16} =
  \mu_p M_{e-}^{p-} m\sqrt{2nh} \:  e_1 e_2 A^-
  I^{(1)}_{l_e, n} I^{(2)}_{n - 1, l_p - 1}.
\]
Here, $e_{1,2}$ are the $z$-components of the photon polarization
vectors, and the following notations are introduced:\vspace*{-1mm}
\begin{equation}
  M_{e\pm}^{p\pm} = \sqrt{\tilde m_e \pm \mu_e m} \sqrt{\tilde m_p \pm \mu_p m},
  \end{equation}\vspace*{-7mm}
\begin{equation}
  T_j^\pm = e_{j x} \pm i e_{j y}, \quad j = 1, 2,
\end{equation}\vspace*{-7mm}
\begin{equation}
  A^\pm = \frac{R_e^2 R_p^2 \pm p_{ez} p_{pz}}{R_e R_p},
  \end{equation}\vspace*{-5mm}
\begin{equation}
  S^\pm = \frac{R_e^2 p_{pz} \pm R_p^2 p_{ez}}{R_e R_p},
\end{equation}
where $R_{e,p}$ are determined by Eq.\,({\ref{Rep}}).\,\,The
functions {$I_{l,l^{\prime}}^{(j)}$} are typical of QED problems
with the magnetic field {\cite{Klepikov54, FIAN}}:\vspace*{-2mm}
\begin{equation}
  I_{l, l'}^{(j)} = e^{-\Phi_j}
  \frac{(i\sqrt{\eta_j})^{\Lambda-\lambda}} {(\Lambda-\lambda)!}
  \sqrt{\frac{\Lambda!}{\lambda!}} F(-\lambda, \Lambda - \lambda +1, \eta_j),
\end{equation}\vspace*{-5mm}
\begin{equation}
  \Lambda = \max (l, l'), \quad \lambda = \min (l, l'),
\end{equation}\vspace*{-7mm}
\begin{equation}
  \Phi_j = \frac{\eta_j}{2} + \frac{k_{jx}\kappa_{j}}{2m^2h} + i\phi_j (l - l'),
\end{equation}\vspace*{-6mm}
\begin{equation}
  \kappa_1 = 2p_{ey} - k_{1y},
\end{equation}\vspace*{-7mm}
\begin{equation}
  \kappa_2 = k_{2y} - 2p_{py},
\end{equation}\vspace*{-7mm}
\begin{equation}
  \eta_j =  \frac{k_{jx}^2 + k_{jy}^2}{2m^2h},
\end{equation}
where {$\phi_{j}$} are the azimuthal angles of photons, and
{$F(a,b,\eta)$} is the degenerate hypergeometric function.\,\,The
term that corresponds to the exchange diagram is obtained by
swapping the subscripts: $1\rightleftarrows2$.

\section{Resonance Conditions}

The process acquires a resonant character if its kinematics permits
an on-shell virtual electron state, i.e.\,\,if the quantities
$g_{0}$ and $g_{z}$ satisfy the standard relation between the
electron energy and the momentum in a magnetic field,
\begin{equation}
\label{rescond}
  g_0^2 = g_z^2 + m^2 + 2nhm^2 = E_n^2.
\end{equation}

In this case, the denominator in propagator (\ref{G}) vanishes. In
order to avoid the resulting divergence, the radiation width of
a virtual state, $\Gamma
$, is introduced according to the Breit--Wigner prescription \cite{Graziani95},%
\begin{equation}
  \label{Breit}
  E_n \rightarrow E_n - \frac{i}{2} \Gamma.
\end{equation}

Let us determine the resonance frequencies of photons in the
ultra-quantum approximation corresponding to the conditions
\begin{equation}
  l \sim 1, \quad lh \ll 1.
\end{equation}
Note that those conditions are characteristic of subcritical
fields.\,\,We confine the consideration to the case where both
photons propagate normally to the magnetic field, and their energy
is close to the threshold one, i.e.
\begin{equation}
\label{bias}
  \begin{array}{l}
    \displaystyle k_{1z} = k_{2z} = 0, \\[3mm]
 \displaystyle W =  \tilde{m}_e + \tilde{m}_p  + \delta W,  \quad \delta W \lesssim mh.
  \end{array}
\end{equation}
Then, expression (\ref{pz}) for the momentum simplifies,
\begin{equation}
  p_{ez} = \sqrt{m\delta W}.
\end{equation}

Taking into account that the conservation laws are satisfied at the
diagram vertices, we use condition (\ref{rescond}) to find the
resonance frequencies
for the first diagram,%
\begin{equation}
\label{resfreq}
  \begin{array}{l}
   \displaystyle  \omega_1 = mh (l_e - n), \\[3mm]
   \displaystyle \omega_2 = 2m + mh(l_p + n) + \delta W.
  \end{array}
\end{equation}

From whence, it is evident that the pair is produced by the hard
photon, whereas the soft one stimulates the transition of a virtual
electron between the Landau levels.\,\,This result is not
unexpected, because the process at the resonance can be regarded as
the sequence of a one-photon pair production and a photon
absorption.

\section{Resonant Cross-Section}

Let us estimate the resonant cross-section of the two-photon pair
production in the ultra-quantum approximation.\,\,Conditions
(\ref{bias})--(\ref{resfreq}) are supposed to be satisfied.\,\,In
addition, let us select the energetically favorable particle
polarizations \cite{Novak08, Novak09} and the numbers of Landau
levels that correspond to the lowest resonance:\vspace*{-2mm}
\begin{equation}
  \label{favor}
  \mu_e = -1, \quad \mu_p = +1,
\end{equation}\vspace*{-9mm}
\begin{equation}
 \label{lowres}
  l_e = 1, \quad l_p = n =0.
\end{equation}

In the sum over $n$ in amplitude (\ref{Sfi1}), we keep only the
resonant term with $n=0$, in which the denominator tends to
zero.\,\,In this case, the second diagram does not give a
substantial contribution to the process, because the resonance
conditions for it are not obeyed.

Taking the aforesaid, as well as expressions for resonance
frequencies (\ref{resfreq}) and conditions (\ref{favor}) and
(\ref{lowres}), into account, let us expand amplitude (\ref{Sfi1})
in a series in the small parameter $h$.\,\,In the zero-order
approximation, we obtain
\[
 S_{fi} \approx  e^{i\Phi}e^2 (2\pi)^4
  \frac{e^{-\frac 1h }\sin \alpha_1 \cos \alpha_2}{SV (p_0^2 -
  E_n^2)}\,\times
\]\vspace*{-6mm}
\begin{equation}
\times\,
  \delta^3(\gamma_1 + \gamma_2 - e^- - e^+).
\end{equation}

According to the known QED rules, the process cross-section is determined as
the squared probability amplitude multiplied by the interval of final states
and divided by the flux $j$, i.e.%
\begin{equation}
\label{gencross}
  d\sigma = \frac{|S_{fi}|^2 }{j T} \frac{Sd^2p_e}{(2\pi)^2} \frac{Sd^2p_p}{(2\pi)^2},
\end{equation}\vspace*{-7mm}
  \begin{equation}
  j = (1 - \cos \chi)/V,
\end{equation}
where $d^{2}p_{e(p)}=dp_{e(p)y}dp_{e(p)z}$, $T$ is the time, and
$\chi$ the angle between the photons.

The integration of Eq.\,(\ref{gencross}) over $d^{2}p_{p}$ can be
executed with the help of $\delta$-functions.\,\,The process
cross-section does not depend on $p_{ey}$.\,\,Therefore, the
integration over $dp_{ey}$ is reduced to the multiplication of the
expression by $p_{ey}$.\,\,Hence, there emerges a multiplier
$Sp_{ey}/V$.\,\,According to work \cite{Klepikov54}, we can get rid
of it by identifying the normalization length $L=V/S$ with the
coordinate of the Larmor orbit center,
$x_{0}=p_{ey}/m^{2}h$.\,\,Hence,
\begin{equation}
  \frac{Sp_{ey}}{V} = m^2 h.
\end{equation}
The remaining integral over $dp_{ez}$ can be found using the
relation
\begin{equation}
  \delta (\omega_1 + \omega_2 - E_e - E_p) =
  \frac{m}{2p_{ez}} \sum \delta(p_{ez} \pm \sqrt{m\delta W}).
\end{equation}

In addition, let us express the quantities $\alpha_{1}$ and $\alpha_{2}$
in terms of the more convenient Stokes parameters,
\begin{equation}
\begin{array}{l}
 \displaystyle  \sin^2 \alpha_j  = \frac{1}{2} (1 -
 \xi_3^{(j)}),\\[3mm]
  \displaystyle \cos^2 \alpha_j  = \frac{1}{2} (1 + \xi_3^{(j)}).
\end{array}
\end{equation}

The ultimate expression for the resonant cross-section looks like
\begin{equation}
  \label{final}
  \sigma = \alpha^2 \frac{\pi h e^{-\frac{2}{h}}}{8(1 - \cos\chi)}
                      \sqrt{\frac{m}{\delta W}}
                      \frac{(1 - \xi_3^{(1)})(1 + \xi_3^{(2)})}{ (g_0 - E_n)^2  + \frac{\Gamma^2}{4} }\!,
\end{equation}
where $\alpha$ is the fine structure constant.

\section{Conclusions}

Let us analyze the obtained expression (\ref{final}) and
de\-mon\-stra\-te that the resonant cross-section can be factorized
into the probabilities of the first-order pro\-ces\-ses.\,\,In the
ultra-quantum approximation, the probabilities of the one-photon
$e^{-}e^{+}$ pair production and the magnetobremsstrahlung look like
\cite{Novak09}%
\begin{equation}
  \frac{dW^{\rm rad}_{1,0}}{du} \bigg|_{u=0} = \alpha \frac{mh^2}{4} (1 - \xi_3^{(1)}),
\end{equation}\vspace*{-5mm}
\begin{equation}
  W^{pp}_{1,0} = \alpha \frac{m h}{4\sqrt{\delta W/m}} e^{-\frac{2}{h}} (1 + \xi_3^{(2)})
\end{equation}
(here, conditions (\ref{rescond}), (\ref{favor}), and (\ref{lowres})
were used). Hence, the process cross-section (\ref{final}) can be
written in the form
of Breit--Wigner formula,%
\begin{equation}
  \sigma = \frac{2\pi}{m^2h^2 (1 - \cos\chi)}
            \frac{ \frac{dW^{\rm rad}_{1,0}}{du} |_{u=0} W^{pp}_{1,0}}{ (g_0 - E_n)^2 + \frac{\Gamma^2}{4}}.
\end{equation}

At last, let us evaluate cross-section (\ref{final}) in the case of
the head-on collision of photons ($\chi=\pi$).\,\,As an example, we
choose the following
 values of parameters:
\begin{equation}
  \xi_3^{(1)} = -1, \quad
  \xi_3^{(2)} =  1,
\end{equation}\vspace*{-7mm}
  \begin{equation}
  h = 0.1, \quad
  \delta W = mh.
\end{equation}
Then the radiation width and the cross-section approximately equal
\begin{equation}
  \Gamma \approx 7 \times 10^{17} \: \mathrm{s}^{-1}\!,
\end{equation}\vspace*{-9mm}
\begin{equation}
  \sigma \approx 2 \times 10^{-2} \:\mathrm{barn}.
\end{equation}

\vspace*{-5mm}
\rezume{%
М.М.\,Дяченко, О.П.\,Новак, Р.І.\,Холодов}{ПОРОГОВЕ РЕЗОНАНСНЕ
ДВОФОТОННЕ\\ НАРОДЖЕННЯ $e^-e^+$ ПАРИ В СИЛЬНОМУ\\ МАГНІТНОМУ ПОЛІ
НА НАЙНИЖЧІ РІВНІ ЛАНДАУ} {В роботі розглянуто процес народження
електрон-по\-зит\-рон\-ної  пари двома фотонами в сильному
магнітному полі. Досліджена кінематика та знайдена загальна
амплітуда процесу з довільною поляризацією частинок. Знайдено умови
резонансного перебігу реакції та проведено оцінку перерізу для
випадку, коли електрон та позитрон займають найнижчі рівні ($l_e
=$~1, $l_p =$~0), що задовольняють умови резонансу.}


\noindent

\vspace*{-2mm}
\begin{thebibliography}{99}                                                                                               %


\bibitem {Baur02}G.~Baur, K.~Hencken, D.~Trautmann \textit{et al}., Phys.
Rep. \textbf{364}, 359 (2002).\vspace*{-0.3mm}

\bibitem {Baur07}G.~Baur, K.~Hencken, and D.~Trautmann, Phys.
Rep. \textbf{453}, 1 (2007).\vspace*{-0.3mm}

\bibitem {Greiner}W.~Greiner, B.~M\"{u}ller, and J.~Rafelski, \textit{Quantum
Electrodynamics of Strong Fields} (Springer, Berlin,
1985).\vspace*{-0.3mm}

\bibitem {CDR}W.~Henning, \textit{FAIR Conceptual Design Report} (Ge\-sell\-schaft
f\"{u}r Schwerionenforschung, Darmstadt, 2001).\vspace*{-0.3mm}

\bibitem {Soff81}G.~Soff, J.~Reinhardt, and W. Greiner, Phys. Rev. A
\textbf{23}, 701 (1981).\vspace*{-0.3mm}

\bibitem {Rumrich87}K.~Rumrich, W.~Greiner, and G.~Soff, Phys. Lett. A
\textbf{125}, 394 (1987).\vspace*{-0.3mm}

\bibitem {Soff88}G.~Soff and J.~Reinhardt, Phys. Lett. B \textbf{211}, 179 (1988).\vspace*{-0.3mm}

\bibitem {Fomin98}P.I.~Fomin and R.I.~Kholodov, Dopov. Nat. Akad. Nauk Ukr.,
No. 12, 91 (1998).\vspace*{-0.3mm}

\bibitem {Backe78}H.~Backe, L.~Handschug, F.~Hessberger \textit{et al}., Phys. Rev.
Lett. \textbf{40}, 1443 (1978).\vspace*{-0.3mm}

\bibitem {Koenig87}W.~Koenig, F.~Bosch, P.~Kienle \textit{et al}., Z. Phys. A
\textbf{328}, 129 (1987).\vspace*{-0.3mm}

\bibitem {Cowan86}T.~Cowan, H.~Backe, K.~Bethge \textit{et al}., Phys. Rev. Lett.
\textbf{56}, 444 (1986).\vspace*{-0.3mm}

\bibitem {LandauIV}V.B.~Berestetskii, E.M.~Lifshitz, and L.P.~Pitaevskii,
\textit{Relativistic Quantum Theory} (Pergamon Press, Oxford,
1982).\vspace*{-0.3mm}

\bibitem {Budker78}G.I.~Budker and A.N.~Skrinskii, Usp. Fiz. Nauk
\textbf{124}, 561 (1978).\vspace*{-0.3mm}

\bibitem {Meshkov94}I.N.~Meshkov, Elem. Chast. At. Yadro \textbf{25}, 1487 (1994).\vspace*{-0.3mm}

\bibitem {Parkhomchuk00}V.V.~Parkhomchuk and A.N.~Skrinskii, Usp. Fiz. Nauk
\textbf{170}, 473 (2000).\vspace*{-0.3mm}

\bibitem {Dikanskii88}N.S.~Dikanskii, N.Kh.~Kot, V.I.~Kudelainen \textit{et al.}, Zh.
\`{E}ksp. Teor. Fiz. \textbf{94}, 65 (1988).

\bibitem {Galnander09}B.~Galnander \textit{et al.}, \textit{HESR Electron Cooler Design
study. Technical report} (Svedberg Laboratory, Uppsala University, Uppsala, 2009).

\bibitem {Bazhenov03}O.~Bazhenov \textit{et al}., \textit{Electron Cooling for HESR.
Final Report} (G.I. Budker Institute of Nuclear Physics,
Novosibirsk, 2003).

\bibitem {Larkin59}A.I.~Larkin, Zh. \`{E}ksp. Teor. Fiz. \textbf{37}, 264 (1959).

\bibitem {Akhiezer61}I.A.~Akhiezer, Zh. \`{E}ksp. Teor. Fiz. \textbf{40}, 954 (1961).

\bibitem {Ng77}Y.~Ng and W.~Tsai, Phys. Rev.~D. \textbf{16}, 286 (1977).

\bibitem {Kozlenkov86}A.A.~Kozlenkov and I.G.~Mitrofanov, Zh. \`{E}ksp. Teor.
Fiz. \textbf{91}, 1978 (1986).

\bibitem {Fomin00}P.I.~Fomin and R.I.~Kholodov, Zh. \`{E}ksp. Teor. Fiz.
\textbf{117}, 319 (2000).

\bibitem {Fomin99}P.I.~Fomin and R.I.~Kholodov, Ukr. Fiz. Zh. \textbf{44},
1526 (1999).

\bibitem {Graziani95}C.~Graziani, A.K.~Harding, and R.~Sina, Phys. Rev. D
\textbf{51}, 7097 (1995).

\bibitem {Novak08}O.P.~Novak and R.I.~Kholodov, Ukr. J. Phys. \textbf{53}, 185 (2008).

\bibitem {Novak09}O.P.~Novak and R.I.~Kholodov, Phys. Rev. D \textbf{80},
025025 (2009).

\bibitem {Klepikov54}N.P.~Klepikov, Zh. \`{E}ksp. Teor. Fiz. \textbf{26}, 19 (1954).

\bibitem {FIAN}V.I.~Ritus and A.I.~Nikishov, Trudy Fiz. Inst. Akad. Nauk SSSR
\textbf{111}, 1 (1979).
\begin{flushright}
{\footnotesize Received 20.02.14.\\ Translated from Ukrainian by
O.I.~Voitenko}
\end{flushright}
\end{thebibliography}
\end{document}